# The Classical Schrödinger's equation

**Bogdan Mielnik**[a,b]
**and**
**Marco A. Reyes**[a]

[a]*Departamento de Física*
*Centro de Investigación y Estudios Avanzados del IPN*
*Apdo. Postal 14-740, México D.F. 07000, México*

[b]*Department of Physics*
*Warsaw University, Warsaw, Polland*

# The Classical Schrödinger's Equation

Bogdan Mielnik[a,b]
and
Marco A. Reyes[a]

[a]*Departamento de Física*
*Centro de Investigación y Estudios Avanzados del IPN*
*Apdo. Postal 14-740, México D.F. 07000, México*

[b]*Department of Physics*
*Warsaw University, Warsaw, Polland*

**Abstract**

A non perturbative numerical method for determining the discrete spectra is deduced from the classical analogue of the Schrödinger's equation. The energy eigenvalues coincide with the bifurcation parameters for the classical orbits.

# 1 Introduction

One of known curiosities of the one–dimensional Schrödinger's equation is its reduciblity to the 1–st order differential equation of Riccati. Given the eigenvalue problem for the Schrödinger's wave packet $\psi(x)$:

$$-\frac{1}{2}\frac{d^2}{dx^2}\psi(x) + [V(x) - E]\psi(x) = 0 \tag{1}$$

(where $m = \hbar = 1$), the formal substitution:

$$\psi(x) = e^{\phi(x)} \tag{2}$$

leads to the classical Riccati equation for $f(x) = \phi'(x)$:

$$f'(x) + f(x)^2 = 2[V(x) - E] \tag{3}$$

Despite its apparent simplicity, (3) is one of non–trivial and persistently returning problems in mathematical physics. In fact, even the field equations of the General Relativity might be viewed as a 'spiritual analogue' of the Riccati eq. (see e.g. [1]). An obvious idea would be to replace completely the spectral problem (1) by its 1–st order equivalent (3). The difficulties, though, are formidable. In the first place, the ansatz (2) is impractical if $\psi(x)$ is a real sign changing function. Moreover, (3) has a tendency to create singularities of $f(x)$, which appear even for $V(x) = const$, and correspond to the nodal points of $\psi(x)$ $[V(x) \equiv V_0 \;\; \Rightarrow \;\; f(x) = -k \;\tan(kx)$, where $k = \sqrt{2(E-V_0)}]$. The subtler ansatz of Aharonov and Au [2] liquidates only a part of this difficulty.

A different chapter in these attempts was open by the angular analog of the Riccati eq.(3) found by Drukarev [3] and Franchetti [4]. Yet, their works appeared in wrong time (wrong place?) and did not receive the attention which they deserved. Drukarev and Franchetti themselves, apparently, have formulated their equations just to facilitate the calculation of the phase shifts (which they did, but the achievement later faded together with the whole "phase shift trend"). As a result, the multiple forms of the Riccati eq. remain a kind of sophisticated curiousm, basically known but scarsely applied. Notable exceptions are the techniques of evaluating the number of energy levels for one dimensional Schrödinger's operator developed in [4, 5, 6, 7, 8], and specially works of Calogero [9, 10]. Yet, the capacities of the method to determine the energy spectra are far from explored. Our purpose here is to complete the story, by studying the exact numerical consequencies of the angular equations [3, 4]. We shall show, that their numerical integration is, most likely, one of the most powerful approximate methods to determine the eigenvalues of one–dimensional (or spherical) spectral problems, which might reduce the traditional perturbation calculus, at least for 1–dimensional eigenvalue problems, into a kind of museal relict.

## 2 The classical model of the Schrödinger's equation.

Curiously, the meaning of our algorithm is best seen by forgetting completely about the quantum mechanical sense of the Schrödinger's equation and sticking to its classical equivalent. This point of view, though seldom applied, has some notable traditions (see, e.g. [11]). Its most provocative expression was the description of the Saturn rings as the band spectrum of the Schrödinger's operator [12].

To perceive the classical sense of (1) denote the variable $x$ by $t$ and call it *time*; put also q$= \psi(t)$, p$= \psi'(t)$. Equation (1) then becomes:

$$\frac{d\mathrm{q}}{dt} = \mathrm{p}, \qquad \frac{d\mathrm{p}}{dt} = 2[V(t) - E]\mathrm{q} \tag{4}$$

Note, that (4) coincide with the canonical equations for the pair of classical variables q,p defined by a time dependent Hamiltonian:

$$H(t) = \frac{\mathrm{p}^2}{2} + [E - V(t)]\mathrm{q}^2 \tag{5}$$

representing the classical oscillator with a time dependent quadratic potential $V(\mathrm{q}, t) = [E - V(t)]\mathrm{q}^2$. The classical phase space trajectory of (4-5):

$$\mathbf{q}(t) = \left\| \begin{matrix} \mathrm{q}(t) \\ \mathrm{p}(t) \end{matrix} \right\| \qquad (t \in \mathbf{R}) \tag{6}$$

'paints' a detailed image of the Schrödinger's wave function $\psi(x)$ and its first derivative $\psi'(x)$. This fact has been decisive to represent the spectral bands of the Schrödinger's operator as the stability bands of a classical system (the Saturn's rings! [12]). It will be as decisive in our description of point spectra.

## 3 The bifurcations.

Assume for simplicity, that none of $E$, $V(t)$ is positive, and $V(t)$ vanishes outside of a finite interval $[a,b] \in \mathbf{R}$ ($V(t)=0$ for $t<a$ and $t>b$). The equations (4-6) thus describe a classical point moving under the influence of a constant repulsive potential $E\mathrm{q}^2$, corrected by an attractive term $-V(t)\mathrm{q}^2$. For $E<0$, the typical trajectory tends to $\pm\infty$ as $t \rightarrow \pm\infty$ (for great $|t|$ the repulsive force dominates, expulsing the classical point to infinity). Occasionally, however, a curious dynamical phenomenon occurs: some trajectories, emerging from the phase space origin (q=p=0) at $t=-\infty$, by a rare coincidence, acquire a momentum sufficient exactly to return asimptotically to the origin, against the repulsive forces. This phenomenon, extremally unstable, represents precisely the eigenvectors of (1) [$\psi(x) \rightarrow 0$ as $x \rightarrow \pm\infty$], i.e. the most stable forms of quantum motion. The effect, in a sense, looks as a classical skill game (whose "goal" is to collocate the material point at the repulsion center; see [13]). To explain the mechanism of the game, some geometry elements on the phase plane $\mathcal{P}$ are needed.

Due to the linearity of (4-5) the phase vector $\mathbf{q}(t)$ at any $t \in \mathbf{R}$ depends linearly on the initial vector $\mathbf{q}(a)$ for any $a \in \mathbf{R}$, i.e.:

$$\mathbf{q}(t) = u(t,a)\mathbf{q}(a), \tag{7}$$

where the $2\times 2$ simplectic real matrix $u(t,a)$ is called the *transference* or *evolution matrix*. The eqs. (4) traduce themselves easily into the first order matrix equation for $u(t,a)$:

$$\frac{du}{dt} = \Lambda(t)u(t,a) \tag{8}$$

where:

$$\Lambda(t) = \left\| \begin{array}{cc} 0 & 1 \\ 2[V(t)-E] & 0 \end{array} \right\| \tag{9}$$

Mathematically, the matrix equation (8–9), is neither easier nor more difficult to solve than the original Schrödinger's equation (1). Yet, it leads to geometric pictures which facilitate the solution of the spectral problem (1).

Consider first of all the region outside of the potential well, $\Omega = \mathbf{R}\backslash(a,b)$, where $V(t)\equiv 0$. The generator $\Lambda(t)$ in $\Omega$ is constant:

$$\Lambda(t) \equiv \Lambda = \left\| \begin{array}{cc} 0 & 1 \\ 2\,|E| & 0 \end{array} \right\| \quad \text{for } t \leq a \text{ or } t \geq b, \tag{10}$$

leading to the explicit solutions:

$$\mathbf{q}(t) = \begin{cases} e^{\Lambda(t-a)}\mathbf{q}(a) & \text{for } t \leq a \\ e^{\Lambda(t-b)}\mathbf{q}(b) & \text{for } t \geq b \end{cases} \tag{11}$$

Note, that $\Lambda$ fulfills:

$$\Lambda^2 = 2\,|E|\,\mathbf{1} \tag{12}$$

and so, it has two eigenvalues $\lambda_\pm = \pm\sqrt{2\,|E|}$ and eigenvectors $\mathbf{e}_\pm$:

$$
\begin{array}{ll}
\textit{eigenvalues} & \textit{eigenvectors} \\
\lambda_+ = +\sqrt{2\,|E|} & \mathbf{e}_+ = \left\| \begin{array}{c} 1 \\ +\sqrt{2\,|E|} \end{array} \right\| \\
\lambda_- = -\sqrt{2\,|E|} & \mathbf{e}_- = \left\| \begin{array}{c} 1 \\ -\sqrt{2\,|E|} \end{array} \right\|
\end{array}
\tag{13}
$$

Henceforth, for $t \in \Omega$ (i.e. in absence of $V(t)$) the motion on the phase plane $\mathcal{P}$ preserves the directions $\mathbf{e}_\pm$ producing a continuous squeezing: the direction $\mathbf{e}_+$ expands while $\mathbf{e}_-$ exponentially shrinks as $t \to +\infty$ (inversely for $t \to -\infty$). (The squeezing is typically generated by the hamiltonians of repulsive oscillators, while the attractive ones generate the phase space rotations).

The phase trajectory (6-7), in general, diverges in both $+\infty$ and $-\infty$. However, exceptions exist. If the initial phase vector $\mathbf{q}(a)$ is proportional to $\mathbf{e}_+$, then in agreement with (11) $\mathbf{q}(t)$ vanishes as $exp[(t-a)\sqrt{2\,|E|}]$ for $t \to -\infty$. In turn, if $\mathbf{q}(b)$ is proportional to $\mathbf{e}_-$, then $\mathbf{q}(t)$ vanishes as $exp[-(t-b)\sqrt{2\,|E|}]$ for $t \to +\infty$. A number $E < 0$ is an *eigenvalue* of the Schrödinger's operator in (1) *iff* there exists a non–trivial trajectory vanishing *on both extremes* $t \to -\infty$ and $t \to +\infty$. This can happen if and only if the evolution between $t = a$ and $t = b$ brings $\mathbf{e}_+$ into a vector proportional to $\mathbf{e}_-$, i.e:

$$
u(b,a)\mathbf{e}_+ = Const \times \mathbf{e}_- \tag{14}
$$

To monitor the phenomenon, consider the integral trajectories of (4-5) for different values of $E < 0$ [or alternatively, for different amplitudes of $V(t)$]. Let $I_+, I_-, -I_+, -I_-$ be the radial $\frac{1}{2}$–lines sticking from the phase space origin in the directions of $\mathbf{e}_+$,$\mathbf{e}_-$,$-\mathbf{e}_+$, $-\mathbf{e}_-$ and let $\mathcal{P}^+$ and $\mathcal{P}^-$ denote two open *connected components* of $\mathcal{P}\backslash I_-$ containing $I_+$ and $-I_+$ respectively (see Fig.1). In absence of $V(t)$, $\mathcal{P}^+$ and $\mathcal{P}^-$ are evolution invariant; as $t \to +\infty$, all the trajectories in $\mathcal{P}^+$ escape to infinity tending asimptotically to $I_+$–axis while all the trajectories in $\mathcal{P}^-$ tend to $-I_+$. Consider now the trajectory which departs from $\mathbf{q}(-\infty) = \mathbf{0}$ crossing $\mathbf{q}(a) = \mathbf{e}_+(E)$. If $V \equiv 0$, $\mathbf{q}(t) = exp[(t-a)\sqrt{2\,|E|}]\mathbf{e}_+$ (i.e., our trajectory escapes to infinity exactly along the $I_+$–axis). If $|V(t)|$ is non zero but small in $[a,b]$, the escape along $I_+$ (due to the squeezing/expansion) is corrected by $-V(t)\mathrm{q}^2$ which generates a clock–wise rotation in $\mathcal{P}$. Yet, once the rotation dies out at $t \geq b$, the squeezing takes over and drives the trajectory back again to $I_+$ as $t \to +\infty$. If $|V(t)|$ is stronger [or alternatively, if $|E|$ is weaker] the trajectory will deviate more from $I_+$ and return to it slower (Fig.1). Finally, for sufficiently small $|E|$ (or sufficiently strong $|V|$), the rotation caused by $V(t)$ will bring $\mathbf{q}(b)$ to $I_-$ (see Fig.1). Instead of returning to $I_+$, the phase point will fall down right to the origin along the (shrinking) direction $I_-$, drawing a (normalized) eigenvector of (1). For $|E|$ still smaller, $\mathbf{q}(b)$ will cross to $\mathcal{P}^-$ and the trajectory will alter its asimptotic behaviour: it will tend now to $-I_+$ as $t \to +\infty$. For $|E|$ further decreasing [or $|V|$ increasing], $\mathbf{q}(b)$ will circulate on $\mathcal{P}$ crossing subsequently $-I_-$, $I_-$, $-I_-$, $I_-, \ldots$. Every such event will produce a closed trajectory (an eigenvector!), coinciding with an abrupt change of the asymptotic type. The eigenvalues of the energy operator are thus interpretable as *bifurcation values* (i.e.

the values of $E$ for which the classical orbits produce bifurcations) [An idea arises, that the discrete spectrum of (1) could be simply *defined in terms of bifurcations*. Such definition would not require linearity, and so, could be easily extended to the non-linear Schrödinger's operators (see the discussions in [14, 15] ].

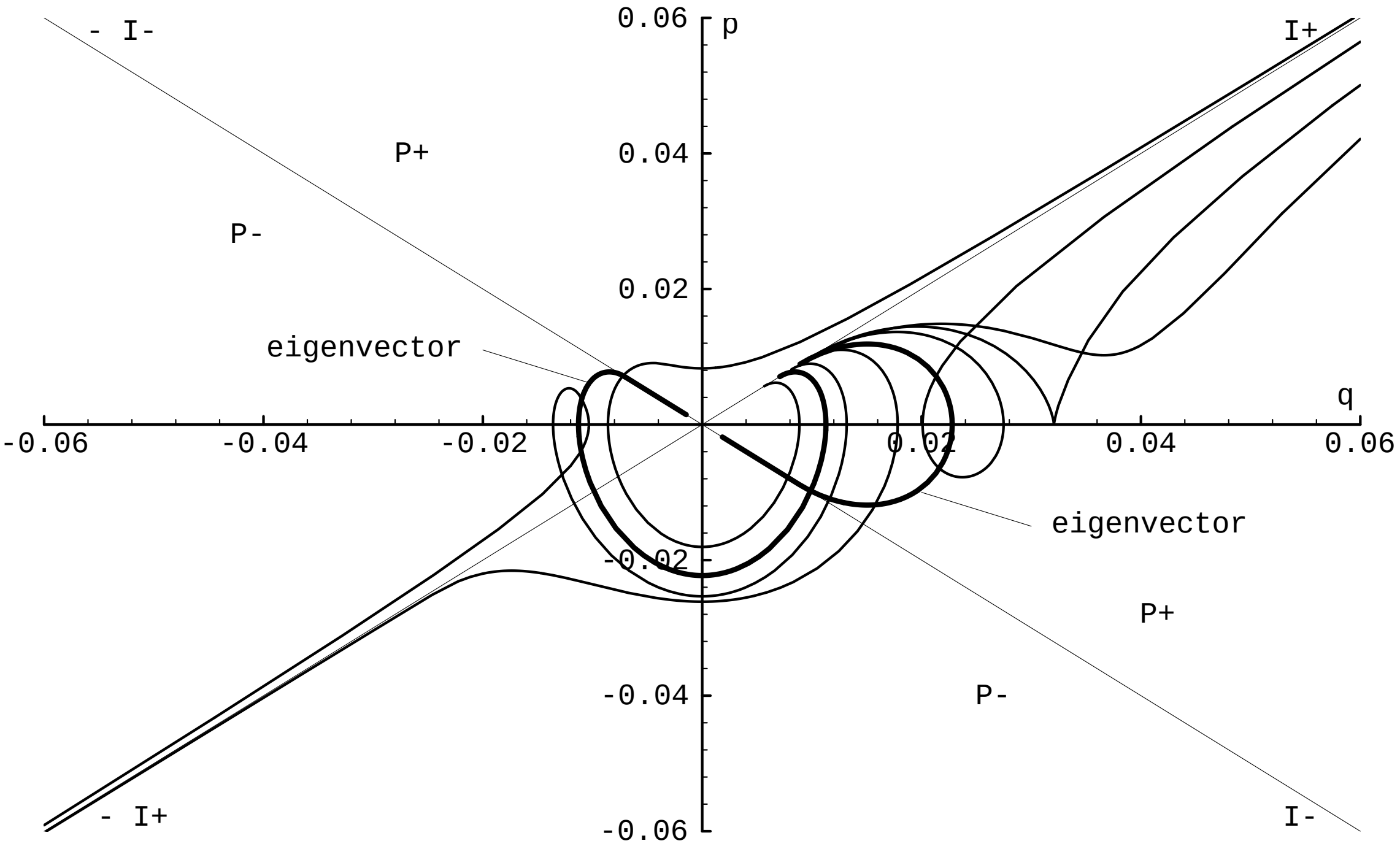

**Figure 1.** The metamorphosis of the classical trajectory (4–5) for varying $E$ and fixed $V(t)$ (qualitative picture). As $E$ raises to zero from below, the deformation due to the rotating term $-V(t)\mathrm{q}^2$ expands clockwisely around the phase space origin, crossing several times the "shrinking axis" $\mathbf{e}_-(E)$. At each new intersection a bifurcation occurs, producing a new closed orbit interpretable as an eigenvector of the Schrödinger's equation (1). The trajectory transformations are pictured in the moving frame of the 'squeezing axis' and represent as well the bifurcations which must occur for a fixed $E < 0$ and variable $V(t)$.

What not less interesting, the *bifurcation values* can be determined by a *rather* simple algorithm, *independent of the traditional perturbation methods.*

# 4 The angular Riccati equations.

Since the vector norms in (14) are irrelevant, (14) can be conveniently written in terms of an *angular coordinate*. Indeed, introduce:

$$\mathrm{q} = \rho\cos\alpha, \quad \mathrm{p} = \rho\sin\alpha \tag{15}$$

The canonical equations become:

$$\dot\rho\cos\alpha - \dot\alpha\rho\sin\alpha = \rho\sin\alpha \tag{16}$$
$$\dot\rho\sin\alpha + \dot\alpha\rho\cos\alpha = 2[V(t) - E]\rho\cos\alpha \tag{17}$$

where $\dot\rho$ and $\dot\alpha$ mean the time derivatives. Curiously, the equation for the angular variable separates. Multiplying (16) by $-\sin\alpha$, (17) by $\cos\alpha$ (or viceversa) and adding one gets the 1–st order differential equation for $\alpha$ alone:

$$\dot\alpha = 2[V(t) - E]\cos^2\alpha - \sin^2\alpha \tag{18}$$

and simultaneously:

$$\dot\rho/\rho = [V(t) - E + 1/2]\sin 2\alpha \tag{19}$$

The equation (18) was first derived by Drukarev [3] and Franchetti [4] (by using purely algebraic arguments). Its link to the original Ricatti idea is immediate. In fact, (18) implies:

$$\frac{d}{dt}\tan\alpha + \tan^2\alpha = 2[V(t) - E] \tag{20}$$

The advantages of the Drukarev-Franchetti form over the traditional Riccati equation (3), however, are that: 1) the eq. (18) can be solved for arbitrary $\alpha$'s without leading to singularities; 2) it offers a clear geometric idea of the spectral condition, and 3) it leads to an easy numerical algorithm. Indeed, notice, that the phase vectors $\mathbf{e}_\pm$ define the angles:

$$\alpha_\pm(E) = \pm\arctan\sqrt{2\,|E|} \tag{21}$$

$$\delta(E) = \alpha_+(E) - \alpha_-(E) = 2\arctan\sqrt{2\,|E|} \tag{22}$$

In agreement with (11), the trajectory tends to zero in $-\infty$ iff $\alpha(a) = \alpha_+ + k\pi$, and it tends to zero in $+\infty$ iff $\alpha(b) = \alpha_- + m\pi$. The trajectory tends to zero for both $t \to \pm\infty$ iff the time evolution (18) converts the initial angle $\alpha(a) = \alpha_+$ into $\alpha(b) = \alpha_- \pm n\pi$. Thus, to check whether a number $E < 0$ is an energy eigenvalue it is sufficient to solve the differential equation (18) with $E$ fixed and with the initial condition $\alpha(a) = \alpha_+(E)$, finding $\alpha(b) = \alpha(b, E)$. Whenever the total angular change $\Delta\alpha = \alpha(a) - \alpha(b)$ in $[a, b]$ fulfills:

$$\Delta\alpha = \delta(E) + n\pi, \qquad n = 0, 1, 2, \ldots \tag{23}$$

$E$ belongs to the discrete spectrum. A convenient form of (23) involves the *spectral defect angle* defined as a difference between the *desired angle* $\alpha_-(E)$ and the *achieved angle* $\alpha(b, E)$:

$$\Gamma(E) = \alpha_-(E) - \alpha(b, E) = \Delta\alpha - \delta(E) \tag{24}$$

The condition (23) then tells:

$$\Gamma(E) = n\pi \qquad n = 0, 1, 2, \ldots \tag{25}$$

An immediate generalization of the conditions (23–25) is obtained for $V(t)$ constant but not necessarily vanishing outside of $(a, b)$:

$$V(t) = \begin{cases} V(a) & \text{for } t \leq a \\ V(b) & \text{for } t \geq b \end{cases} \tag{26}$$

The motion (4-6) has then two different constant generators $\Lambda(a)$ and $\Lambda(b)$ for $t \leq a$ and $t \geq b$ and the formulae (23–25) hold after substituting $|V(a) - E|$ or $|V(b) - E|$ instead of $|E|$ in the expression (21) for $\alpha_+$ and $\alpha_-$ respectively.

Observe, that (15) is not the only way to separate the angular part of (4). The $\mathcal{P}$-plane is a simplectic space without a natural measure of distances and angles. Henceforth, instead of introducing the polar variable (15) straighforwardly, one might as well introduce new canonical coordinates q',p', and only after define the angular variable on q',p' plane. Some profits of this freedom were explored by Calogero [9]; here let us notice only the plausible form of the spectral condition if the angular variable is introduced by:

$$\begin{aligned} \mathrm{q} &= \lambda \rho \cos \alpha \\ \mathrm{p} &= \lambda^{-1} \rho \sin \alpha \end{aligned} \tag{27}$$

where $\lambda = (2\,|E|)^{-\frac{1}{4}}$. The motion equations then read:

$$\dot{\alpha} = \sqrt{2\,|E|}\, \cos 2\alpha + \sqrt{2/\,|E|}\, V(t) \cos^2 \alpha \tag{28}$$

$$\dot{\rho}/\rho = (\sqrt{2\,|E|} + V(t)\sqrt{|E|\,/2}\,) \sin 2\alpha \tag{29}$$

and the limiting angles are independent of $E$ ($\alpha_\pm = \pm\frac{\pi}{4}$). The spectral condition becomes:

$$\Delta\alpha = (n + \frac{1}{2})\pi \tag{30}$$

in a visible reconciliation between the oscillator spectrum and the Sommerfeld quantization conditions.

# 5 The numerical algorithm.

The importance of the angular variables for the structure of the bound states was detected already in papers devoted to the phase shifts [17]. The *defect angle* equivalent to (24) was introduced indeed by Calogero, and used to evaluate the number of the energy levels for the 1–dimensional and radial Schrödinger's eigenproblem [10]. Yet, the efficiency of the method to determine the exact eigenvalues somehow escaped attention (perhaps, due to a general fascination by the perturbative methods!). The main reason, why (18) is so easily applicable, is:

**Theorem 1.** For any bounded, piece–wise continuous $V(t)$ with a compact support in $[a, b]$, $\Gamma(E)$ is a strictly increasing function of $E$ for $E < V(a)$.

The **proof** is deduced from two observations. (1) If $\alpha(t)$ and $\alpha'(t)$ are two solutions of (18) with $\alpha(a) < \alpha'(a)$ then $\alpha(b) < \alpha'(b)$ [Indeed, otherwise there would be a point $c \in [a, b]$ with $\alpha(c) < \alpha'(c)$ contradicting the uniqueness of the solutions of (18)]; (2) If $\alpha_1(t)$ and $\alpha_2(t)$ are solutions of two equations of form (18) with two different parameters $E = E_1$ and $E = E_2$ respectively, then $\alpha_1(a) = \alpha_2(a)$, and $E_1 > E_2$ (and) $\alpha_1(a) < \alpha_2(a)$ [the proof involves only the elementary analysis of the Cauchy eq.(18)]. The observations (1) and (2) imply now that $\alpha(b, E)$ for an angular trajectory starting in $\alpha(a) = \alpha_+(E)$ is a monotonically decreasing function of $E$, and the proof is completed by noticing that $\alpha_-(E)$ is increasing.

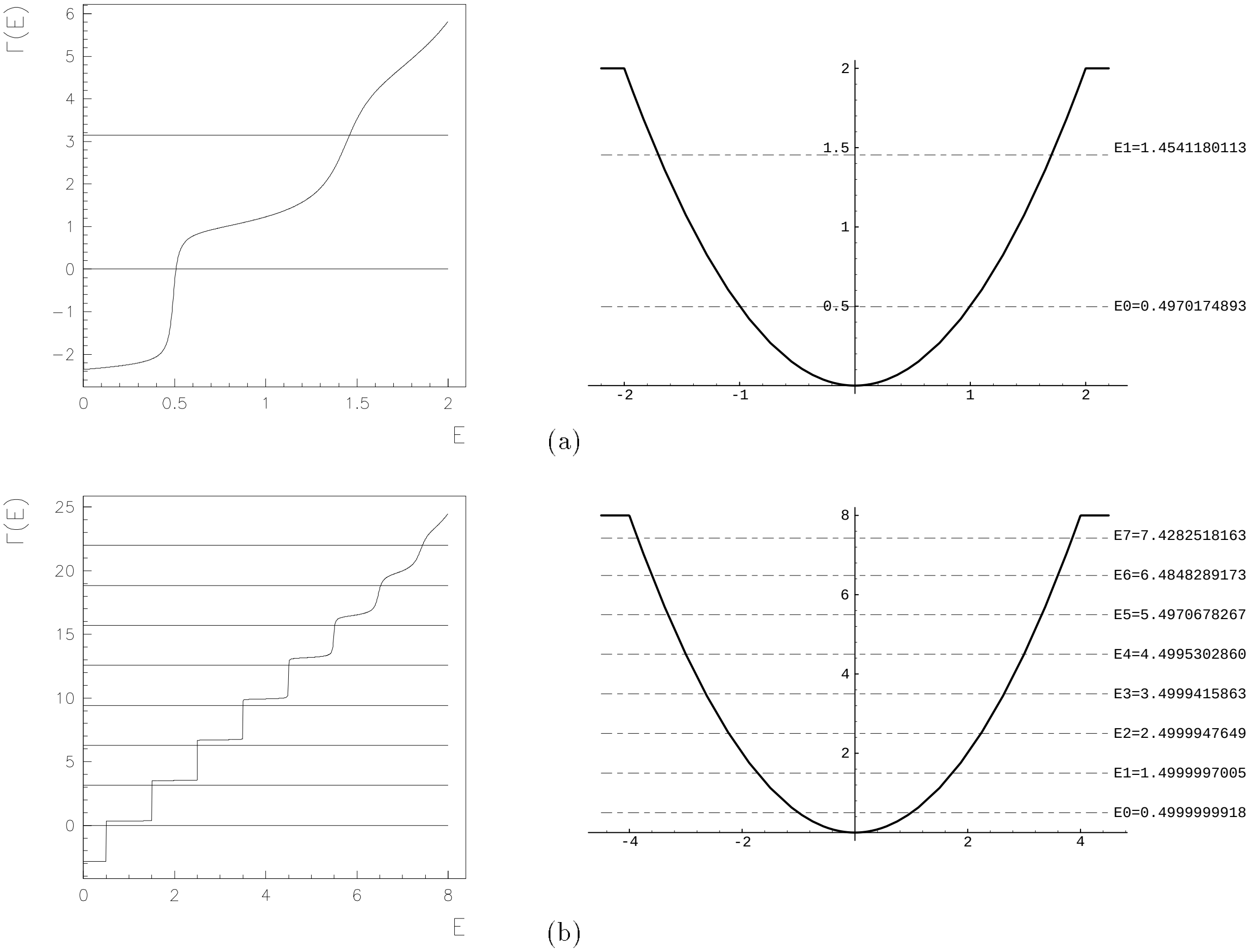

**Figure 2.** The defect angle $\Gamma(E) = \alpha_- - \alpha(a, E)$ for two cases of truncated oscillators: (a) $b = -a = 2$ and (b) $b = -a = 4$. The intersections of the "stepping" functions $\Gamma(E)$ with the lines $n \times \pi$ give the eigenvalues of the Schrödinger problem.

As an illustration, we have used (25) to determine the energy levels for the truncated 1–dimensional oscillator potential:

$$V(x) = \begin{cases} \frac{1}{2}\omega^2 x^2 & \text{for } |x| \leq a \\ \frac{1}{2}\omega^2 a^2 & \text{for } |x| \geq a \end{cases} \tag{31}$$

The limiting angles are $\alpha_\pm = \pm \arctan \sqrt{w^2a^2 - 2E}$. We have determined the angular function $\Gamma(E)$, $0 < E < V(a)$, for $w = 1$, $a = 2$, and for $a = 4$, by integrating numerically the angular equation (18) (see Fig.2). It yields the 2 energy levels for the oscillator truncated at $a = 2$, and 8 energy levels for the oscillator truncated at $a = 4$, all calculated with accuracy up to $10^{-10}$. Curiously, the obtained eigenvalues are very close to the first 2 and 8 levels of the exact oscillator, respectively, $E_n = n + \frac{1}{2}$ (indeed, even the last eigenvalues of the truncated potential (31) differ rather little from the orthodox $E_1 = 1.5$, and $E_7 = 7.5$). Comparing to Ritz method, a basic advantage of our algorithm is its essential simplicity (no need to waste skills inventing an adequate class of the test functions). A notable advantage is that the spectral function $\Gamma(E)$ is *unstable* and changes very quickly when crossing the sequence of critical values $\Gamma = n\pi$ $(n = 1, 2, \ldots)$ (see Fig.2). Thus, even a very little error in $E$ in the vecinity of an eigenvalue, traduces itself into a visible effect in $\Gamma$, significantly improving the accuracy (compare Calogero [16, p 274]). This instability is caused by the fact that the energy eigenvalues correspond to the bifurcations of the orbits and the final point $\alpha(b)$ deflects very fast when $E$ crosses the bifurcation value. In limit $b = +\infty$, $\alpha(b, E)$ would be discontinuous and $\Gamma(E)$ would be an exact step function!

## 6 The "$\frac{1}{2}$ eigenproblem".

Our method, till now, concerns only the potentials constant outside of finite intervals (limited, non–singular wells). Could it tell something about more general $V(t)$? Consider any continuous $V : \mathbf{R} \to \mathbf{R}$ such that:

$$V_\pm = \liminf_{t \to \pm\infty} V(t) > -\infty \tag{32}$$

In the traditional approach to the spectral problem, the main effort is to find the non–trivial solutions of (1) vanishing on both extremes $t \to \pm\infty$ (which exists only as an exception!) Following the observations of Sec.5 we propose to reduce the solution to two minor steps, each one interpretable as "$\frac{1}{2}$ spectral problem": 1) for any $E$ find the special solutions of (1) which vanish for $t \to -\infty$ (the "left eigenvectors"); 2) find the solutions of (1) which vanish for $t \to +\infty$ (the "right eigenvectors").

While the difficulty of solving the complete spectral problem is formidable, one seldom pays attention to the fact, that every "half of it" has a solution *always*, defining some *left (right) decaying branches* for (1) for any $E < V_-$ $(E < V_+)$. For big $|t|$ they provide the *asimptotic cues* $\mathbf{e}_\pm(t, E)$ and *asimptotic angles* $\alpha_\pm(t, E)$ adequate to replace the fixed vectors $\mathbf{e}_\pm(E)$ (13) and *angles* $\alpha_\pm(E)$ in the algorithm of Sec.5. Indeed, one has:

**Lemma** Let $\beta : \mathbf{R} \to \mathbf{R}$ be a continuous real function with:

$$\liminf_{t \to +\infty} \beta(t) > \eta^2 > 0 \tag{33}$$

Then the 2–dimensional solution space $\Xi$ of the 2nd–order differential equation:

$$-\frac{d^2 \mathrm{q}}{dt^2} = \beta(t)\mathrm{q}(t) \tag{34}$$

has a 1–dimensional subspace $\Xi_-$ of solutions which vanish for $t \to +\infty$ and are square integrable in $[0, +\infty)$.

**Proof.** Let $N$ be a number such that $\beta(t) > \eta^2$ for $t \geq N$. For $t \geq N$, the material point $\mathrm{q}(t)$ moves under the influence of the repulsive ellastic force

$$\ddot{\mathrm{q}} \geq \eta^2 \mathrm{q} \tag{35}$$

Consider now an integral trajectory of (33) which satisfies the initial condition: $\mathrm{q}(N) = C > 0$, $\dot{\mathrm{q}}(N) = \eta C > 0$. Using (35) one easily shows that $\mathrm{q}(t)$ is positive, monotonically increasing and:

$$\mathrm{q}(t) \geq \mathrm{q}(N) e^{\eta(t-N)} = K e^{\eta t} \tag{36}$$

The method of variation of constant then provides a new, linearly independent solution:

$$\mathrm{q}_-(t) = \mathrm{q}(t) \int_t^{+\infty} \frac{d\tau}{\mathrm{q}(\tau)^2} \leq \mathrm{q}(t) \int_t^{+\infty} \frac{d\tau}{\mathrm{q}(t)\mathrm{q}(\tau)} = \int_t^{+\infty} \frac{d\tau}{\mathrm{q}(\tau)} \leq \frac{e^{-\eta t}}{C\eta} \tag{37}$$

which spans the desired subspace $\Xi_-$.□

An immediate consequence is:

**Theorem 2.** Let $V(t)$ be a continuous potential in the 1–dimensional Schrödinger's equation (1) and assume (32) holds. Then for every $E < V_-$, the 2–dimensional solution space of (1) contains a 1–dimensional subspace $\Xi_+(E)$ of solutions vanishing for $t \to -\infty$, square integrable in $(-\infty, 0]$, while for any $E < V_+$ it contains a 1–dimensional subspace $\Xi_-(E)$ of solutions vanishing for $t \to +\infty$, square integrable in $[0, +\infty)$. A number $E <\min(V_-, V_+)$ is a point of the discrete spectrum of (1) if the subspaces $\Xi_-(E)$ and $\Xi_+(E)$ coincide.

**Proof** is obtained by applying the **Lemma** to $t$ and $-t$ with $\beta(t) = V(t) - E$ and $0 < \eta^2 < V_\pm - E$ respectively.

In the Appendix we have collected the asimptotic forms of both "decaying branches" of (1) for several $V(t)$ including the oscillator and Coulomb potentials (see Appendix). These cues (the "$\frac{1}{2}$–eigenvectors"), permit to solve, with any desired accuracy, the traditional spectral problem for a class of infinite potential wells. Let $V(t)$ be one of these potentials and let $E_0 < V_\infty$. The energy eigenvalues $E < E_0$ (if they exist), can be then determined by the following algorithm, generalizing the one of Sec.5:

i) one fixes a finite interval $[a, b] \subseteq \mathbf{R}$ such that $V(t) - E_0 \geq \kappa > 0$ and the asimptotic cue expressions be valid for $t \leq a$ and $t \leq b$;

ii) one uses the "vanishing cues" $q_\pm(t, E)$ to define two special angles:

$$\alpha_-(a, E) = \arctan[p_-(a, E)/q_-(a, E)] \tag{38}$$

$$\alpha_+(b, E) = \arctan[p_+(b, E)/q_+(b, E)] \tag{39}$$

characteristic for the trajectories vanishing at $t \to -\infty$ and $t \to +\infty$.

iii) one integrates the equation (18) in $[a, b]$ with the initial condition $\alpha(a) = \alpha_+(a, E)$ finding $\alpha(b) = \alpha(b, E)$ and determining the *defect angle*:

$$\Gamma(E) = \alpha_-(b, E) - \alpha(b, E) \tag{40}$$

iv) whenever:

$$\Gamma(E) = n\pi, \qquad n = 0, 1, 2, \ldots \tag{41}$$

the number $E < E_0$ belongs to the point spectrum of the Schrödinger's operator.

The non–trivial part of the method, of course, is to find the vanishing cues [18]. However, as the interval $[a, b]$ can be arbitrarily wide, it is enough to know the asimptotic expressions. Now, if (32) holds and $E < E_0$, the cues are monotonic, without zeros for $t < a$ and $t > b$; one can thus use the anzatz (2–3). Note that while the general solution $f$ of the Riccati equation (3) depends on one arbitrary constant, and (tipically) diverges as $t \to \pm\infty$, the $f(t)$ of *vanishing cue* has no such arbitrariness, and can be determined with any desired accuracy by applying the known iterative or asymptotic methods (see Appendix). For the oscillator and Coulomb potentials the cues ("$\frac{1}{2}$–eigenvectors) are already known in form of the asimptotic series for the confluent hypergeometric equation, vanishing either at $t \to 0_+$ or $t \to \infty$ (however, we prefer to represent them in the form $\mathrm{q}(t) = exp\{f(t)\}$, since $f(t) = \tan \alpha$ defines the asimptotic angles).

What quite essential, once the cues $\mathrm{q}_\pm$ (and the angles $\alpha_\pm$) are determined, they can be used not only to find the spectrum for one particular potential $V(t)$ but simultaneously, for an entire class of potentials which share the asimptotical behaviour of $V(t)$ (and can be arbitrarily deformed in any finite region). Moreover, given the "left cues" for one potential $V_1(t)$ and the "right cues" for another potential $V_2(t)$ the method can be as well used to determine the spectrum of any $V(t)$ sharing the asymptotic behaviour of $V_1(t)$ for $t \to -\infty$ and of $V_2(t)$ for $t \to +\infty$.

For curiosity, we have used the asimptotic angles calculated in our Appendix to determine the spectrum for the "hybrid oscillator" (Fig.3) not so easily treatable by either perturbative or variational methods:

$$V(x) = \begin{cases} \frac{1}{2}x^2 & \text{for } x \geq 0 \\ \\ \frac{1}{8}x^2 & \text{for } x \leq 0 \end{cases} \tag{42}$$

Note, that our method permits to solve as easily any other hybrid or deformed cases (as e.g. an oscillator affected by an arbitrarily high potential barriere in the middle, etc.)

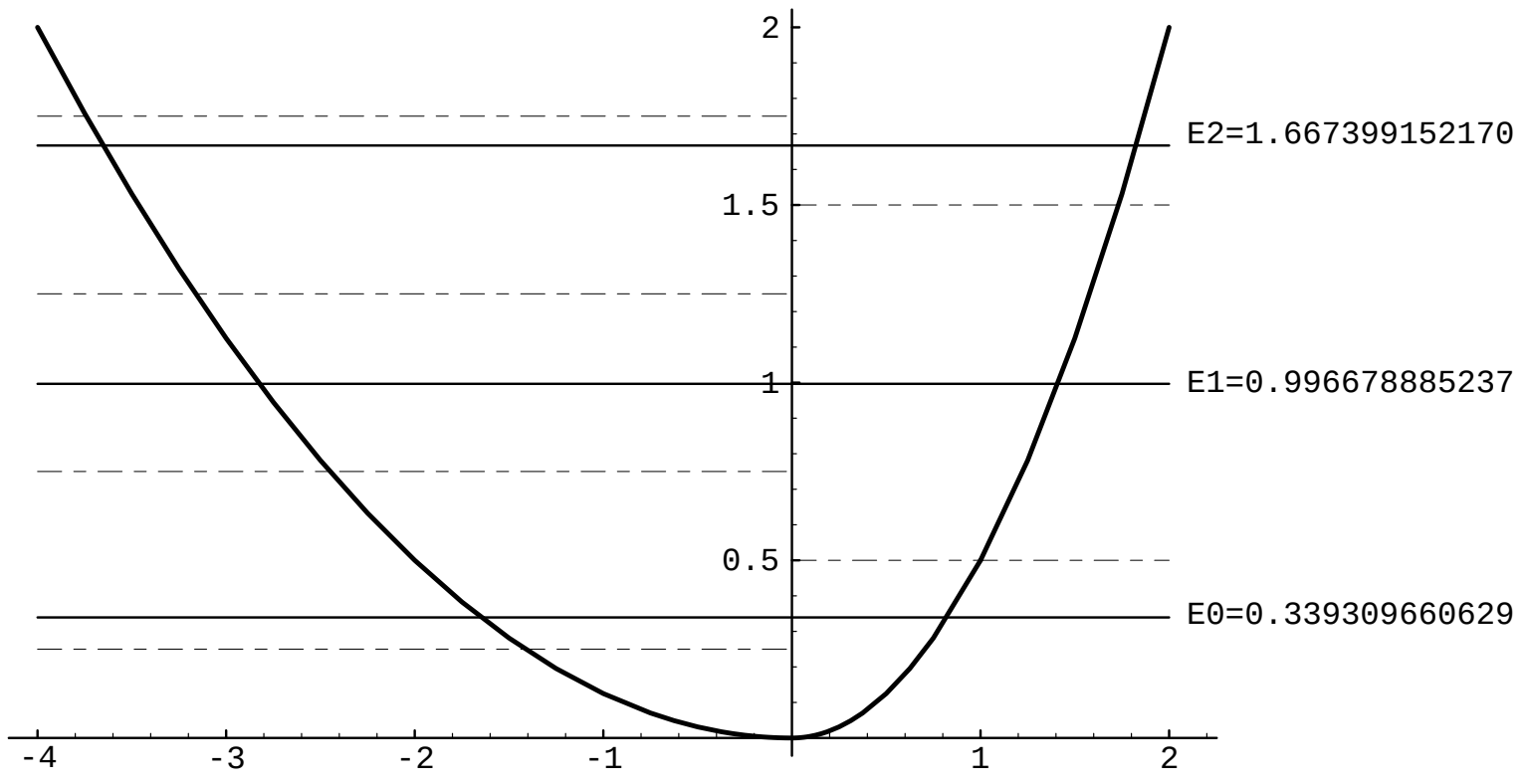

**Figure 3.** The hybrid oscillator potential. The dashed lines represent the standard energy eigenvalues for each separated oscillator, and the solid lines represent the actual eigenvalues for this potential.

# 7 Singular and radial wells.

The physically important wells not only extend to infinity, but can have singularities in the finite region. The typical case is the 1–dimensional eqs. (1) obtained after the separation of the angular variables in the Schrödinger's eq. in $\mathbf{R}^3$ with a radial potential $\phi(r)$. By denoting $\psi(\mathbf{x}) = R(r)Y(\theta,\phi)$ and assuming $Y(\theta,\phi)$ to be an eigenfunction of the square angular momentum $\mathbf{L}^2$ , one ends up with the 1–dimensional eigenvalue problem for $u(t) = tR(t)$:

$$-\frac{d^2u}{dt^2} + [2V(t) + \frac{l(l+1)}{t^2} - 2E]u = 0 \tag{43}$$

whose eigenvectors are the trajectories vanishing for $t \to 0_+$ and square integrable in any $[b, +\infty)$ $(b > 0)$. Similarly as before, each of the asimptotical conditions, typically, can be satisfied by solutions of (1) for any $E$. The existence of the *right cues* (solutions vanishing for $t \to +\infty$) is assured by Theorem 2. In turn, the existence of the *zero-cues* (solutions vanishing at $0_+$) is the consequence of the following elementary theorem:

**Theorem 3.** Let $\phi(t)$ be a continuous real function in $(0, +\infty)$, satisfying:

$$0 \neq \phi(t) \geq -k(t) \tag{44}$$

in a certain subinterval $(0, \delta > 0)$, where $k : (0, \delta) \to \mathbf{R}$ fulfills:

$$k(t) \geq 0 \quad \text{and} \quad \int_0^\delta \int_0^t k(t')dt' < +\infty \tag{45}$$

Then the 2–dimensional space $\Xi$ of the functions $u : (0, +\infty) \to \mathbf{R}$ which solve the 2nd–order differential equation:

$$-\frac{d^2u}{dt^2}+\phi(t)u=0 \tag{46}$$

must contain a 1–dimensional subspace $\Xi_0$ of solutions which vanish for $t \to 0_+$.

**Proof.** We shall stick to the classical image of Sec.2 putting $u(t)$ =q and $\dot{u}(t)$ =p; (46) then paints the motion of a classical point mass $m=1$ under the influence of the ellastic force $-2\phi(t)\mathrm{q}(t)$. Choose now $a \in (0,\delta)$ and consider the integral trajectories of (46) defined by the initial condition $\mathrm{p}(a)=\dot{\mathrm{q}}(a)=0$ ($\mathrm{q}(a)$ arbitrary). The assumption (45) permits one to evaluate the lower bound of $\mathrm{q}(t)$ in $(0,a)$. Indeed,

$$\begin{aligned}
\ddot{\mathrm{q}}(t) &\geq -k(t)\mathrm{q}(t) \Rightarrow & (47)\\
\Rightarrow \dot{\mathrm{q}}(a)-\dot{\mathrm{q}}(t)=\int_t^a \ddot{\mathrm{q}}(t')dt' &\geq -\int_t^a k(t')\mathrm{q}(t')dt' \geq -\mathrm{q}(a)\int_t^a k(t')dt' \Rightarrow & (48)\\
\Rightarrow \dot{\mathrm{q}}(t) &\leq \mathrm{q}(a)\int_t^0 k(t')dt' \Rightarrow & (49)\\
\Rightarrow \mathrm{q}(a)-\mathrm{q}(t)=\int_t^a \dot{\mathrm{q}}(t')dt' &\leq \mathrm{q}(a)\int_t^a\int_{t'}^a k(t")dt"dt' \leq K(a)\mathrm{q}(a) \Rightarrow & (50)\\
\Rightarrow \mathrm{q}(t) &\geq [1-K(a)]\mathrm{q}(a), & (51)
\end{aligned}$$

where

$$K(a)=\int_0^a\int_t^a k(t')dt'dt. \tag{52}$$

Henceforth, if $a$ is small enough, so that $K(a)<1$, then $\mathrm{q}(t)$ has a positive lower bound in $(0,a)$. If now $\phi(t)<0$ in $(0,a)$, then $\ddot{\mathrm{q}}(t)<0 \Rightarrow \dot{\mathrm{q}}(t)>0=\dot{\mathrm{q}}(a) \Rightarrow \mathrm{q}(t)$ monotonically increases in $(0,a)$ tending to its positive lower bound at $t \to 0_+$. In turn, if $\phi(t)>0$ in $(0,a)$, then $\dot{\mathrm{q}}(t)<0$ and $\mathrm{q}(t)$ monotonically decreases in $(0,a)$ tending to its (finite or infinite) upper bound in $(0,a)$. In both cases, the standard method of the "variation of constant" provides the new solution of (46):

$$\mathrm{q}_0(t)=\mathrm{q}(t)\int_0^t \frac{d\tau}{q(\tau)^2}=t\cdot h(t)\left(\frac{1}{t}\int_0^t\frac{h(\tau)^2d\tau}{h(t)^2}\right) \tag{53}$$

where $h(t)=1/\mathrm{q}(t)$ is positive, monotonic and bounded in $(0,a)$. Note that the expression $(1/t)\int_0^t h(\tau)d\tau$ is the average value of $h(\tau)^2$ in $(0,t)$. If $\phi(t)$ is positive, and $h(t)$ increasing ($\Leftrightarrow \mathrm{q}(t)$ decreasing) the fraction in the square parenthesis in (53) is $\leq 1$; but if $\phi(t)$ is negative and $h(t)$ decreasing, it tends to 1 from above as $t \to 0_+$. In both cases, for $t$ small enough, there is a constant $N>0$ such that $\mathrm{q}_0(t)<Nt$, and so, $\mathrm{q}_0(t)$ defines the desired subspace $\Xi_0$ of solutions vanishing as $t \to 0_+$.□

In the Appendix we have collected examples of asimptotic cues at $0^+$ for some tipical singularities.

To check the results, we have used the right cues ($t \to +\infty$) and the $0_+$ cues of the Appendix to determine the discrete spectrum for the Coulomb well $V(t)=-\frac{1}{t}$, obtaining the first 10 levels of the hydrogen atom with accuracy to $10^{-10}$:

$E_0$ = -0.5000000000
$E_1$ = -0.1250000000
$E_2$ = -0.0555555555
$E_3$ = -0.0312499999
$E_4$ = -0.0200000000
$E_5$ = -0.0138888888
$E_6$ = -0.0102040816
$E_7$ = -0.0078125000
$E_8$ = -0.0061728395
$E_9$ = -0.0050000000

Once the asimptotic expressions were cross–examined, we have used the already tested right cues of the oscillator and the $0_+$ cues of the Coulomb singularity to find the principal series of energy levels for the "hybrid well":

$$V(r) = -\frac{1}{r} + \frac{1}{2}w^2r^2 \tag{54}$$

sometimes considered as a candidate to describe the quark confinement [19]. Taking $w^2 = 0.00005$ we could observe, that a very weak oscillator potential cancels the condensation of the hydrogen energy levels for $E \to 0^-$, providing a continuous transition to an equally spaced spectrum for $E > -0.01$ (see Fig.4). Let us also notice that a close relative of our method has been succesfully used to study the logarithmic wells [20].

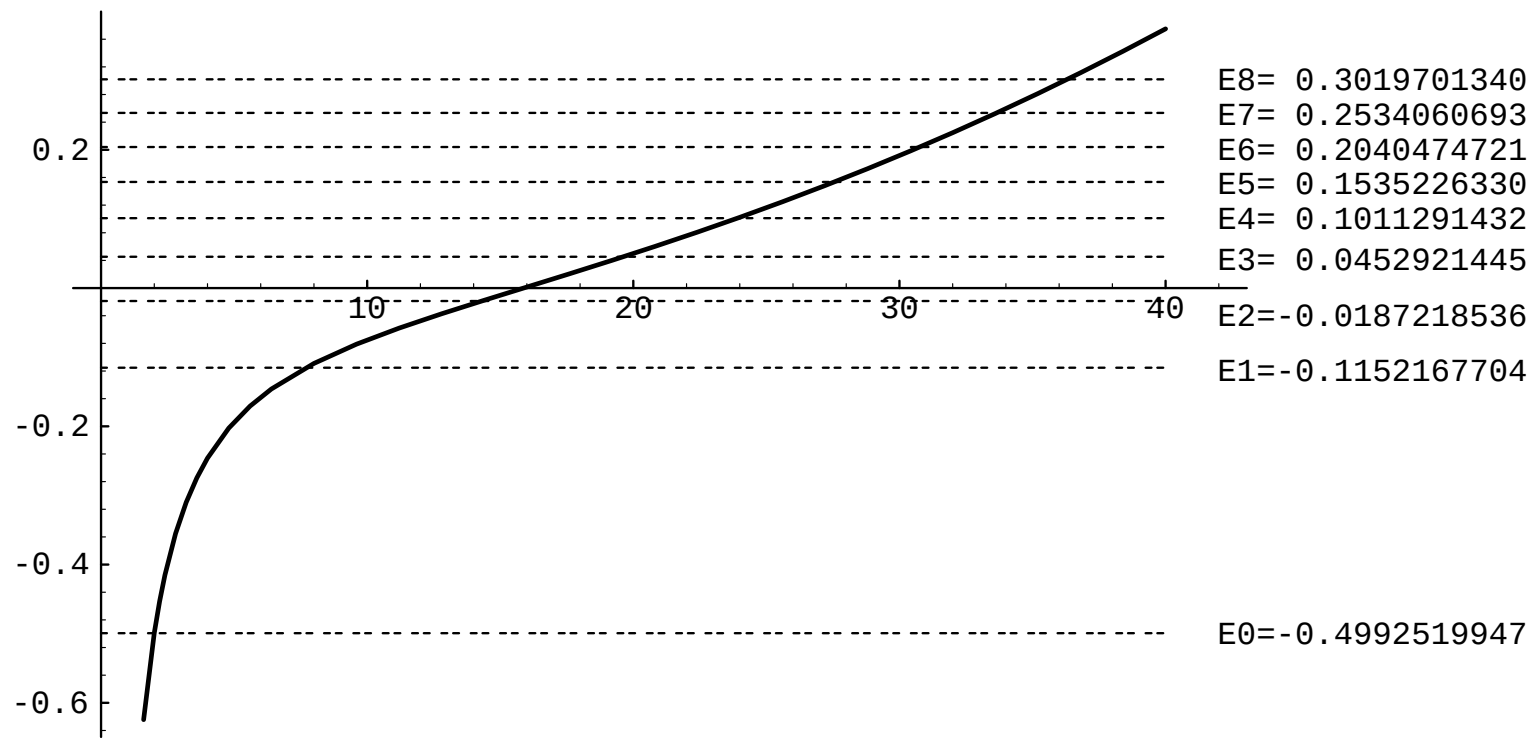

**Figure 4.** A hybrid well: the kind which could resemble the quark confinement.

We are tempted to predict that in not so far future, the perturbative methods of determining spectra (at least for the 1–dimensional Schrödinger's operator) will be almost forgotten and a part of pacience nowadays devoted to special functions (permitting to solve only exceptional problems!) will be invested into building up a bank of data about the asimptotic behaviours and "vanishing cues". Once these data will be precise enough, the task of determining the spectra of arbitrary potentials of known asimptotic types will become a question for the pocket calculators.

# Acknowledgements

The authors are grateful to their Colleagues in Depto. de Física, México, in the Int. Center of Computer Modeling, Warsaw, and in the Institute of Theoretical Physics, Warsaw, for helpful discussions. Thanks are also due to the organizers of the 4th Int. Conf. of Squeezed States and Uncertainty Relations in Shanxi, June 1995, for their interest in the subject and helpful comments. The support of CONACYT, México, is acknowledged.

# Appendix: The evaluation of the vanishing cues

Below, we look straightforwardly for $f = \tan\alpha$ defined by the Riccati equation (3) and yielding $\mathrm{q}(t) = e^{\int f(t')dt'}$ which vanish either for $t \to \pm\infty$ or $t \to 0_+$.

# A Harmonic oscillator

The solution to the Riccati equation

$$\dot{f} + f^2 = w^2 t^2 - 2E$$

which yields $\mathrm{q}(t) \to 0$ for $t \to \pm\infty$, can be found in form of an infinite series:

$$f(t) = -wt + \sum_{i=0}^{+\infty} a_i t^{-i}$$

where

$$a_0 = a_{2s} = 0, \qquad a_1 = \frac{E}{w} - \frac{1}{2},$$

$$a_{s+1} = \frac{1}{2w}\left\{-(s-1)a_{s-1} + \sum_{j=1}^{s} a_j a_{s-j}\right\} \qquad (s \geq 2)$$

Explicitely:

$$\begin{aligned} f(t) \;\; = \;\; & -wt + \left[\frac{E}{w} - \frac{1}{2}\right]\frac{1}{t} + \frac{1}{2w}\left[\frac{E}{w} - \frac{1}{2}\right]\left[\frac{E}{w} - \frac{3}{2}\right]\frac{1}{t^3} + \\ & \frac{1}{2w^2}\left[\frac{E}{w} - \frac{1}{2}\right]\left[\frac{E}{w} - \frac{3}{2}\right]\left[\frac{E}{w} - 2\right]\frac{1}{t^5} + \ldots \end{aligned}$$

The asimptotic form of the "$\frac{1}{2}$–eigenvectors" for $t \to \pm\infty$ is:

$$\mathrm{q}_{\mp} \simeq e^{-\frac{1}{2}wt^2}\; t^{\left(\frac{E}{w} - \frac{1}{2}\right)}\; e^{-2\left(\frac{E}{w} - \frac{3}{2}\right)\frac{1}{t^2} - 4\left(\frac{E}{w} - \frac{3}{2}\right)\left(\frac{E}{w} - 2\right)\frac{1}{t^4} - \ldots}$$

The consistency with the textbooks expressions for $\psi(t)$ in terms of the confluent hypergeometric function is easily verified.

## B The Coulomb potential

The series solution for

$$\dot{f} + f^2 = -\frac{2}{t} + \frac{l(l+1)}{t^2} - 2E$$

[yielding $q_0(t) \to 0$ when $t \to 0_+$] turns out to be

$$f_0 = \frac{l+1}{t} + \sum_{i=0}^{+\infty} a_i t^i$$

Explicitely

$$f_0 = \frac{l+1}{t} - \frac{1}{l+1} - \frac{2E(l+1)^2+1}{(2l+3)(l+1)^2} t + \dots$$

Henceforth, the "$\frac{1}{2}$–eigenvector" vanishing for $t \to 0_+$ is:

$$q_0 \simeq t^{l+1} \; e^{-\frac{1}{l+1}t - \frac{2E(l+1)^2+1}{2(2l+3)(l+1)^2}t^2 + \dots}$$

consistently with the well known expressions in terms of the confluent hypergeometric series.

For $t \to +\infty$, by similar arguments, $f$ has the form

$$f_- = -\sqrt{2|E|} + \frac{1}{\sqrt{2|E|}}\frac{1}{t} - \frac{2l(l+1)|E| + \sqrt{2|E|} - 1}{4|E|\sqrt{2|E|}}\frac{1}{t^2} + \dots$$

yielding the "$\frac{1}{2}$–eigenvector"

$$q_- \simeq e^{-\sqrt{2|E|}t} \; t^{\frac{1}{\sqrt{2|E|}}} \; e^{h(1/t)}$$

where $h(\xi)$ has the form of an anlytic series vanishing at $\xi = 0$. However, we have found that if the integration interval is wide enough, a simpler asimptotic expression gives good results:

$$f_- = -\sqrt{-2E - \frac{2}{t} + \frac{l(l+1)}{t^2}}$$

and the only problem we have is to integrate the angular equation (18) in a sufficiently long range, to have the cue without nodal points. The integration can be simplified by changing variables from $t$ to $x = t/(1+t)$.

## C The Yukawa potential

This case is very similar to the preceding one, though no longer treatable in terms of hypergeometric series. The Riccati equation for $f(t) = \tan\alpha(t)$ reads

$$\dot{f} + f^2 = -\frac{2}{t}e^{-\lambda t} + \frac{l(l+1)}{t^2} - 2E$$

Our Theorem 3 assures the existence of the solution of (1) vanishing at zero, of the form $\psi_0(t) = exp\{f_0(t)\}$ where:

$$f_0 = \frac{l+1}{t} + \sum_{i=0}^{+\infty} a_i t^i$$

After calculations

$$a_0 = \frac{-1}{l+1}, \qquad a_1 = \frac{2(\lambda - E)(l+1)^2 - 1}{(2l+3)(l+1)^2}$$

$$a_{s+1} = \frac{-1}{2l+s+3}\left\{\frac{2(-\lambda)^{s+1}}{(s+1)!} - \sum_{r=0}^{s} a_r a_{s-r}\right\} \qquad (s \geq 1)$$

Henceforth, the vanishing cue $\mathrm{q}_0(t)$ is:

$$\mathrm{q}_0(t) \simeq t^{l+1}\ e^{-\frac{1}{l+1}t + \frac{2(\lambda-E)(l+1)^2-1}{2(2l+3)(l+1)^2}t^2 + \ldots}$$

(no longer representable in terms of the confluent hypergeometric series!)

For $t \to +\infty$, we have used a finite approximant:

$$f_- = \sqrt{\frac{l(l+1)}{t^2} - 2E - \frac{2}{t}e^{-\lambda t}}$$

changing simultaneously the integration variable to $x = t/(1+t)$.

# D A hybrid "quark potential"

The Riccati equation (3) for the effective potential

$$V(t) = -\frac{1}{t} + \frac{1}{2}w^2 t^2 - \frac{l(l+1)}{2t^2}$$

yields the following $f_0(t)$ for $t \to 0_+$:

$$\begin{aligned} f_0 \quad = \quad & \frac{l+1}{t} - \frac{1}{l+1} - \frac{2E(l+1)^2+1}{(2l+3)(l+1)^2}t - \frac{2E(l+1)^2+1}{(2l+3)(l+1)^3(l+2)}t^2 + \\ & \frac{1}{2l+5}\left\{w^2 - \frac{4E(l+1)^2+2}{(2l+3)(l+1)^4(l+2)} - \frac{\left(2E(l+1)^2+1\right)^2}{(2l+3)^2(l+1)^4}\right\}t^3 + \ldots \end{aligned}$$

The vanishing cue

$$\mathrm{q}_0 = t^{l+1}\ e^{-\frac{1}{l+1}t + g(t)}$$

where $g(t)$ has the form of an analytical series.

Meanwhile, the solution vanishing at $t \to +\infty$ corresponds to $f_-(t)$ in the form

$$f_- = -wt + \left[\frac{E}{w} - \frac{1}{2}\right]\frac{1}{t} + \frac{1}{w}\frac{1}{t^2} + \frac{1}{2w}\left\{\left[\frac{E}{w} - \frac{1}{2}\right]\left[\frac{E}{w} - \frac{3}{2}\right] - l(l+1)\right\}\frac{1}{t^3} + \ldots$$

The vanishing cue is

$$\mathrm{q}_- \simeq e^{-\frac{1}{2}wt^2}\ t^{\left(\frac{E}{w} - \frac{1}{2}\right)} e^{k(1/t)}$$

where $k(\xi)$ is an analytic series vanishing at $\xi = 0$.